\documentclass[a4paper,12pt,abstract=true,enabledeprecatedfontcommands]{scrartcl}
\usepackage{authblk}

\usepackage{amsmath, amssymb}
\usepackage{graphicx,color}
\usepackage{bm}
\usepackage[colorlinks=true,allcolors=blue]{hyperref}
\usepackage{cancel}

\allowdisplaybreaks

\begin{document}
\titlehead{\hfill KYUSHU-HET-223, KUNS-2862}

\title{Baryon number non-conservation\\ as Peccei-Quinn mechanism}

\author[1]{Takahiro Ohata}
\affil[1]{Department of Physics, Kyoto University, Kyoto 606--8502, Japan}

\author[2]{Kengo~Takeuchi}
\affil[2]{Department of Physics, Kyushu University, 744 Motooka, Nishi-ku, Fukuoka, 819--0395, Japan}

\author[2]{Koji~Tsumura\thanks{Email: \texttt{tsumura.koji@phys.kyushu-u.ac.jp}}}
          
\maketitle

\begin{abstract}
Baryon number is an accidental symmetry in the standard model, 
while Peccei-Quinn symmetry is hypothetical symmetry which is introduced 
to solve the strong CP problem. 
We study the possible connections between Peccei-Quinn symmetry and 
baryon number symmetry. 
In this framework, an axion is identified as the Nambu-Goldstone boson 
of baryon number violation. 
As a result, characteristic baryon number violating processes are predicted. 
We developed the general method to determine the baryon number and 
lepton number of new scalar in the axion model. 
\end{abstract}

\section{Introduction}
%
%
The last undetermined parameter of the standard model for particle physics 
(SM) is the QCD $\theta$ parameter. 
The $\theta$ parameter is constrained to be extremely small 
by the search for the neutron electric dipole moment\cite{Ref:nEDM}, 
which imply the CP symmetry in the strong interaction. 
However, the CP symmetry is broken in the electroweak sector 
of the SM by the Kobayashi-Maskawa mechanism. 
It is natural to expect the CP violation in the strong interaction 
against non-observation of the strong CP phase $\theta$. 
This unnaturalness is called the strong CP problem\cite{Ref:StrongCP}. 
%

%
%
The strong CP problem may be solved by the hypothetical Peccei-Quinn 
symmetry (PQ symmetry)\cite{Ref:PQ}, 
where the enhanced global symmetry leaves an axion 
as Nambu-Goldstone boson after the symmetry breaking.  
Thanks to the shift symmetry of the axion, 
the QCD $\theta$ parameter becomes unphysical. 
When the axion develops a vacuum expectation value (VEV), 
the vanishing $\theta$ parameter is realized as a physical quantity. 
Thus, the strong CP problem is solved dynamically. 
%
%
The minimal extension of the SM to realize the PQ mechanism is known 
to be ruled out, where the second Higgs doublet is introduced with 
a global symmetry\cite{Ref:WW}.
There are two major axion models. 
One is so-called KSVZ axion\cite{Ref:KSVZ}, 
where the PQ mechanism is realized 
outside the SM sector. 
An SM singlet complex scalar is coupled to the postulated heavy 
colored fermions. 
The other is DFSZ axion model\cite{Ref:DFSZ}, 
where an SM singlet complex scalar 
is added to the original axion model\cite{Ref:WW}.
Assuming the large VEV of the singlet scalar, 
these axion models become invisible against experimental searches. 
%

%
%
Invisible axion models are often combined with the global symmetries, 
which are motivated by the other problem in the SM. 
Froggatt-Nielsen mechanism is a familiar mechanism of the origin of fermion 
mass hierarchy, where a flavor symmetry is assumed. 
If the flavor symmetry and the PQ symmetry are broken by the common VEV, 
then these two independently inspired scenarios have the common physics 
scale and the characterized predictions\cite{Ref:flaxion,Ref:axiflavon}.  
For another example of such model, 
the type-I seesaw mechanism is known to generate 
the lepton number violating Majorana neutrino masses. 
If the lepton number symmetry is identified with PQ symmetry, 
a pseudo Nambu-Goldstone boson called Majoron\cite{Ref:Majoron} plays 
a role of axion\cite{Ref:Shin,Ref:vPQ}.  
%

%
%
The search for the violation of conservation law is believed to be 
a useful probe to access the physics far beyond the TeV scale. 
Although LHC has been reported null evidences of new physics 
below TeV scale, indirect searches using flavor/number as well as CP 
violation have been explored the high energy scale much higher 
than the TeV scale through the virtual mediator effects. 
Non-observation of the nucleon decay search has been provided 
one of the most stringent bound on such process\cite{Ref:n-decay-exp}. 
The searches of the lepton number violating neutrinoless double 
beta decay and the lepton flavor violating processes 
have also been known to be very sensitive probes of new physics. 
%

%
%
In the classical level of the SM, 
the baryon number $\mathbf{B}$ and lepton number $\mathbf{L}$ 
are accidentally conserved. 
However, in the SM effective field theory, these symmetries are, 
in general, violated by the higher dimensional operators. 
For instance, the $\mathbf{L}$ violating operator $LLHH$ is allowed 
at dimension-five, while $\mathbf{B}+\mathbf{L}$ violating operators 
are allowed at dimension-six\cite{Ref:WeinbergOp,Ref:Wilczek}. 
Depending on the operator dimensions, characteristic number violating 
processes are predicted\cite{Ref:WeinbergOp,Ref:Wilczek,Ref:BNV-LNV}.  
If these number violation is identified as PQ symmetry breaking, 
the PQ mechanism may also be explored by the powerful number 
violation searches. 
%

%
%
In this paper, we propose new axion models based 
on the lepton and/or baryon number conservation. 
These number symmetries are identified as the PQ symmetry. 
The strong CP problem is solved by ordinary PQ mechanism, 
while the characteristic number violation is predicted. 
The method to identify the lepton number symmetry by the PQ symmetry 
is developed in Majoraxion models, where the lepton number symmetry 
and PQ symmetry is identified. 
We then generalized the method with the baryon number symmetry. 
As typical examples, we construct axion models which predict 
nucleon decays, $n$-$\bar{n}$ oscillation, and dinucleon decays.  
%

%
%
This paper is organized as follows. 
In Section 2, the Majoraxion models are reviewed and 
formulated by the model independent framework. 
A method to identify the number symmetry by PQ symmetry is developed. 
In Section 3, axion models based on the baryon (and lepton) number violations 
are proposed with typical baryon number violating experimental signatures. 
Conclusion and discussion are given in Section 4. 
%

\section{Anatomy of Majoraxion models}

In this section, we review on known Majoraxion models, and 
then the idea is reformulated in the model independent framework 
using the higher order operators. \\

The idea of the Majoraxion is to unify the axion and the Majoron, 
where the former is a pseudo-Nambu-Goldstone boson (pNGB) induced 
by the PQ symmetry breaking while the latter is one from the 
breaking of the lepton number symmetry. 
The first model\cite{Ref:Shin} of this category is based on the connection 
between the type-I seesaw model and the KSVZ axion model. 
In the type-I seesaw model\cite{Ref:Type-I}, 
the SM singlet right-handed neutrinos $N_R$ 
are postulated to generate the observed neutrino mass and mixing. 
In this model, the lepton number conservation is broken explicitly 
by Majorana mass for right-handed neutrinos. 
The lepton number symmetry can be restored by introducing the 
SM singlet complex scalar $S$ as 
%
\begin{align}
{\mathcal L}_N 
= \Big( -y_N^{}\, \overline{L} \widetilde{H} N_R + \text{H.c.} \Big)
- \frac12 h_N^{}\, S^* \overline{N_R^C} N_R, 
\label{Eq:type-I}
\end{align}
%
where $L$ and $H$ are the lepton and Higgs doublet in the SM, respectively. 
We also note $\widetilde{H}=i\tau_2H^*$ and $\psi^C=C\overline{\psi}^t$. 
Through Yukawa interaction with lepton numberless coupling constant, 
the lepton numbers $\mathbf{L}$ of $N_R$ and $S$ are uniquely determined, 
i.e., $\mathbf{L}(N_R)=+1$ and $\mathbf{L}(S)=+2$. 
In Shin's model\cite{Ref:Shin}, the common complex scalar $S$ is used to break 
the PQ symmetry of KSVZ axion model,\footnote{
The model can also be combined with DFSZ axion model\cite{Ref:nuDFSZ,Ref:nuDFSZ2}. 
As discussed in Ref.\cite{Ref:nuDFSZ2}, $U(1)_\text{PQ}$ is entangled with 
$U(1)_\textbf{B}$, $U(1)_\textbf{L}$ and $U(1)_Y$ in the DFSZ model 
due to the ambiguity of PQ charges of SM fermions. 
On the other hand, PQ symmetry is broken only by SM singlet scalar in the KSVZ model 
resulting no entanglement. Thus, PQ symmetry can be identified as a part of 
$U(1)_\textbf{B} \times U(1)_\textbf{L}$ symmetry. }
%
\begin{align}
{\mathcal L}_\Psi 
= -y_\Psi^{}\, S^*\, \overline{\Psi}_L \Psi_R + \text{H.c.} 
\label{Eq:KSVZ}
\end{align}
%
where $\Psi_L$ and $\Psi_R$ are so-called KSVZ quarks. 
It is now clear that since $\mathbf{L}(\overline{\Psi}_L\Psi_R)\ne0$, 
the lepton number symmetry plays a role of PQ symmetry in this setup. 
Thus, the Majoron is identified as the axion. 
As the simplest choice, $\Psi_L$ and $\Psi_R$ are assumed to transform 
as a fundamental representation of QCD and have single flavor 
in order to easily avoid the domain wall problem\cite{Ref:N_DW=1}. 
If we write the KSVZ Yukawa interaction with $S$ as
%
\begin{align}
{\mathcal L}'_\Psi
= -y_\Psi^{}\, S\, \overline{\Psi}_L \Psi_R + \text{H.c.} 
\label{Eq:KSVZ2}
\end{align}
%
a different lepton numbers assignment is possible. 
\\

Let us reformulate the idea of Majoraxion in a model independent way. 
At renormalizable level, the (global) lepton number symmetry 
is an accidental symmetry of the SM. 
On the other hand, the lepton number is in general broken 
by higher dimensional operators constructed by the SM fields. 
The most popular lepton number violating operator is so-called 
Weinberg operator with mass dimension five\cite{Ref:WeinbergOp}, 
%
\begin{align}
\widehat{\mathcal O}_5 
= LLHH. 
\end{align}
%
This operator breaks lepton number by two units, $\Delta\mathbf{L}=2$. 
What was done in the Majoraxion model is to restore the lepton number 
symmetry by introducing the complex scalar $S$, i.e., 
%
\begin{align}
{\mathcal O}_6
=S^* \widehat{\mathcal O}_5
=S^*LLHH, \label{Eq:S-L}
\end{align}
%
where the lepton number of $S$ is fixed to be $\mathbf{L}(S)=+2$. 
Now, $S$ has a well defined lepton number by Eq.\eqref{Eq:S-L}, 
which is transmitted to $\Psi_L$ and/or $\Psi_R$ through the KSVZ 
Yukawa interaction in Eq.\eqref{Eq:KSVZ2}. 
This prescription is further developed with the baryon number symmetry 
in the next section. 
When $S$ acquires the VEV with this lepton number conserving operator, 
the Majoron appears as an axion. 
If we begin with a seesaw model which has lepton number 
violating dimensionful parameter, the Majoraxion model is 
derived by replacing this operator with the complex scalar $S$. 
That was done in the above example.

\begin{figure}[tb]
\centering 
\includegraphics[scale=0.8]{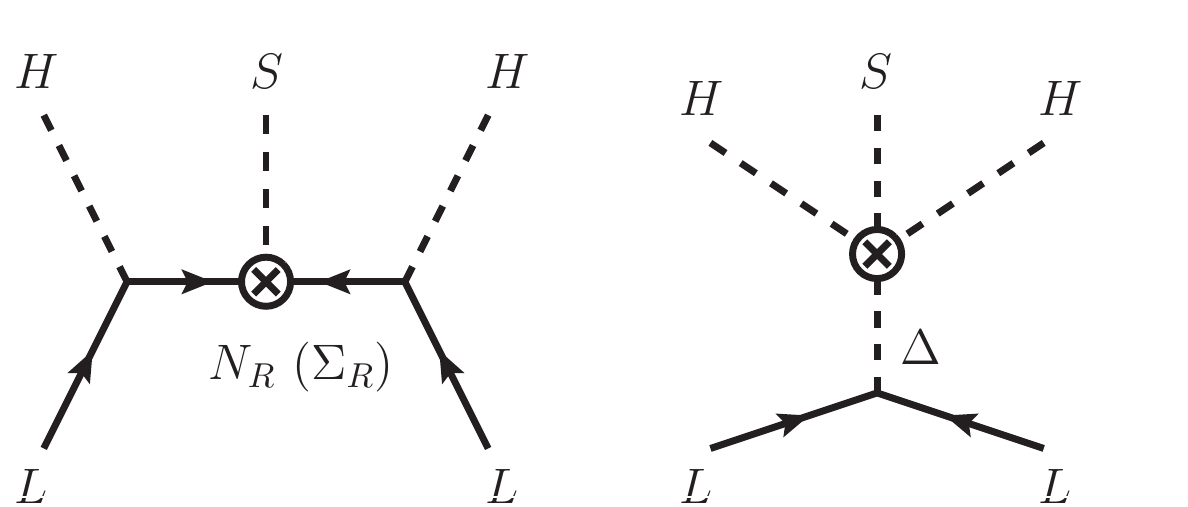}
\caption{Diagrams for tree-level Majoraxion models.}
\label{Fig:Majoraxion_Tree}
\end{figure}

This prescription is applied for the type-II seesaw model\cite{Ref:Type-II}, 
where a complex triplet scalar $\Delta$ with $Y=1$ is introduced.\footnote{
Our hypercharge convention is $Q_\text{EM}=T_3+Y$, where 
the electric charge $Q_\text{EM}$ and the third component of weak isospin $T_3$.}
In the type-II seesaw model, a dimensionful parameter with 
lepton number violation is $\mu$ in the scalar potential, 
$\mathcal{V} \sim \mu\, H^*\Delta H^*$, where the lepton number 
of $\Delta$ is rigorously fixed by the Yukawa interaction 
$y_\Delta^{}L\Delta L$. 
Therefore, by promoting $\mu$ as $S$
we obtain the Majoraxion extension of the type-II seesaw model\cite{Ref:Type-II-Majoraxion}. 
The Majoraxion extension\cite{Ref:Type-III-Majoraxion} 
of type-III seesaw model\cite{Ref:Type-III} is nothing different from 
the type-I seesaw model, where the right-handed neutrino is 
simply replaced by the triplet fermion $\Sigma_R$ with $Y=0$. 
In Fig.~\ref{Fig:Majoraxion_Tree}, the diagrams for $\mathcal{O}_6$ 
are given in type-I (left) and in type-II (right) Majoraxion models. 
The case for type-III seesaw extension is also shown in the left panel. 
If we truncate the external $S$ line in these diagram, 
we obtain the diagrams for the neutrino mass generation 
of lepton number violating dimension five  operator 
$\widehat{\mathcal O}_5$ in ordinary seesaw mechanism. 
\\

\begin{figure}[tb]
\centering 
\includegraphics[scale=0.8]{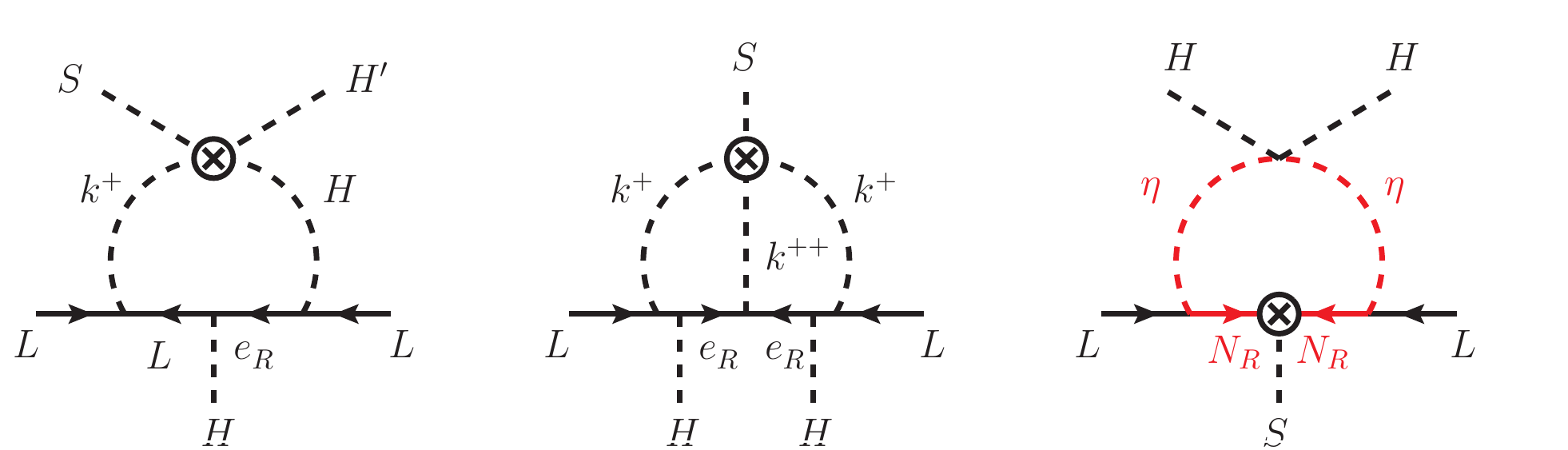}
\caption{Diagrams for loop-level Majoraxion models.}
\label{Fig:Majoraxion_Loop}
\end{figure}

%
The higher dimensional operator $\mathcal{O}_6$ is not necessarily 
decomposed by tree diagrams. 
The above mentioned type-I -II -III seesaw models are based on the 
tree level decomposition\cite{Ref:TreeSeesaw} of the prototype operator 
of $\widehat{O}_5$. 
The loop level classification of $\widehat{O}_5$ is called radiative seesaw model, 
where neutrino masses are generated by quantum corrections. 
The variants of radiative seesaw model have been studied very extensively  
(For a comprehensive review of radiative seesaw models, 
see \cite{Ref:vModelReview}). 
There must be a loop level ultraviolet (UV) completion of $\mathcal{O}_6$, 
that is radiative Majoraxion model\cite{dmt14,maxion}. 
The extension of these radiative seesaw model to the Majoraxion 
model is straightforward. 
In Fig.~\ref{Fig:Majoraxion_Loop}, 
the diagrams for the radiative Majoraxion extensions of 
Zee model (left), Zee-Babu model (center) and scotogenic model (right). 
In ordinary Zee model\cite{Ref:Zee}, a pair of singly charged scalar $k^\pm$ 
and an extra Higgs doublet $H'$ are introduced in order to form 
the lepton number violating connection $\mu_\text{Z}^{}\, k^+H^*H'^*$ 
with dimensionful coupling $\mu_\text{Z}^{}$. 
The substitution of $\mu_\text{Z}^{}$ by $S$ again identifies 
the lepton number symmetry as 
PQ symmetry\cite{Ref:Zee-Majoraxion}. 
In ordinary Zee-Babu model\cite{Ref:Zee-Babu} ,  
a pair of doubly charged scalar $k^{\pm\pm}$ is added instead $H'$, 
then the lepton number violating dimensionful coupling $\mu_\text{ZB}^{}$ 
is allowed as $\mu_\text{ZB}^{}\, k^+k^+k^{--}$.  
By substituting $\mu_\text{Z}^{}$ by $S$, a Majoraxion extension 
of Zee-Babu model is realized\cite{Ref:Type-II-Majoraxion}. 
In the scotogenic model\cite{Ref:Ma}, 
right-handed neutrinos and the so-called inert doublet are assumed 
to be odd under the ad-hoc $Z_2$ symmetry in order to make 
the dark matter stable.  
In the Majoraxion extension of the scotogenic model, 
the dimensionful parameter is provided by the Majorana mass 
of right-handed neutrinos as in type-I seesaw model. 
The stability of the dark matter is automatically guaranteed by the residual 
$Z_2$ symmetry (lepton parity\cite{Ref:DarkParity,Ref:DarkParity2}, 
which is lead by the breakdown of the global lepton number symmetry 
a la Krauss-Wilczek mechanism\cite{Ref:KW}. 
%

\section{$\mathbf{B}$ and $\mathbf{L}$ violation as PQ mechanism}
In this section, we focus on the lepton number and baryon number 
violating operators with mass dimensions more than $d=5$. 
At $d=6$, $\mathbf{B-L}$ conserving operators of $qqq\ell$ 
are allowed, where $q$ and $\ell$ are general quark and lepton fields. 
The $d=7$ operators of $qqq\overline{\ell}\varphi$ hold $\mathbf{B+L}$ 
but violate $\mathbf{B-L}$, where $\varphi$ denotes a SM boson field 
or space-time derivative. 
In this paper, we assume $\varphi$ to be the Higgs field $H$ or 
its charge conjugation. 
For the characterization of $d>7$ baryon/lepton number violating operators, 
see for example\cite{Ref:BNV-LNV}. 

\subsection{Axion models based on $\Delta\mathbf{(B+L)}=2$ symmetry breaking}
The $d=6$ operators of $qqq\ell$ type break both the baryon number and 
lepton number by one unit as $\mathbf{\Delta B}=\mathbf{\Delta L}=1$. 
These are given by 
%
\begin{align}
\widehat{\mathcal O}_6 
= \big\{ u_R^{}u_R^{}d_R^{}e_R^{}, u_R^{}d_R^{}QL, u_R^{}QQe_R^{}, QQQL \big\}, 
\end{align}
%
where SM right-handed quark singlets $u_R^{}$, $d_R^{}$ and 
left-handed quark doublet $Q$, respectively. 
All these operators hold global $\mathbf{B-L}$ symmetry, 
while the $\mathbf{B+L}$ symmetry is explicitly broken. 
According to the prescription discussed in the previous section, 
$\mathbf{B+L}$ symmetry may be restored by introducing 
a complex scalar $S$ as 
%
\begin{align}
{\mathcal O}_7=S^*\,\widehat{\mathcal O}_6. 
\label{Eq:B+L}
\end{align}
%
Note that the baryon and lepton numbers of $S$ are determined as 
$\mathbf{B}(S)=\mathbf{L}(S)=+1$ through this operator. 
In general, the $\mathbf{B+L}$ conserving operator may be 
higher order as ${\mathcal O}_8=(S^*)^2\,\widehat{\mathcal O}_6, \cdots$, 
instead. 
If we commonly use $S^*$ or $S$ for the KSVZ Yukawa interaction 
as in Eq.~\eqref{Eq:KSVZ} or Eq.~\eqref{Eq:KSVZ2}, 
pNGB caused by $\mathbf{B+L}$ symmetry breaking is identified as an axion. 
At this point, how can we assign $\mathbf{B}$ and $\mathbf{L}$ for 
$\Psi_L$ and $\Psi_R$? 

The $\mathbf{B}$ and $\mathbf{L}$ charges for $\Psi_L$ and $\Psi_R$ 
can be determined by introducing the one of the following mass mixing operators, 
%
\begin{align}
{\mathcal O}\,'_\text{mix} 
= \big\{ \overline{Q} \widetilde{H} \Psi_{R}^{U}, \overline{\Psi^U_L}u_R^{}, 
\overline{Q}H\Psi_{R}^{D}, \overline{\Psi^D_L}d_R^{} \big\}. 
\end{align}
%
The hypercharges of $\Psi_L^q$ and $\Psi_R^q$ are chosen to be 
$+2/3$ (for $q=U$) or $-1/3$ (for $q=D$). 
These operators are obtained by replacing a SM quark singlet 
 $q_R^{}(=u_R^{}, d_R^{})$ by $\Psi_{R}^q$ or 
$\overline{Q}\widetilde{H}(\overline{Q}H)$ by 
$\overline{\Psi^U_L}(\overline{\Psi^D_L})$ 
in the quark Yukawa interactions, i.e., $y_U^{}\overline{Q} \widetilde{H} u_R^{}$ 
and $y_D^{}\overline{Q} H d_R$. 
This is usually done in the KSVZ model in order to allow 
the decay of KSVZ quark into the SM particles. 
Through these connections, we find
$\mathbf{B}(\Psi^q_L)=\mathbf{B}(\Psi^q_R)=+1/3$ and 
$\mathbf{L}(\Psi^q_L)=\mathbf{L}(\Psi^q_R)=0$. 
This is not the case for our purpose. 
In order to identify the pNGB as an axion, $\Psi_L$ and $\Psi_R$ must have 
different $\mathbf{B}$ and $\mathbf{L}$ charges. 
Therefore, we determine $\mathbf{B}$ and $\mathbf{L}$ only 
for $\Psi_L^q\,(\Psi_R^q)$ by this procedure, 
and the other link for $\Psi_R^q\,(\Psi_L^q)$ is taken from 
the KSVZ Yukawa interactions through Eq.~\eqref{Eq:B+L}. 
As a concrete example, we assume the operators of $\overline{\Psi^D_L}d_R^{}$ 
and $S^*QQQL$ in our effective Lagrangian. 
At this point, $\mathbf{B}$ and $\mathbf{L}$ for $\Psi^D_L$ and $S$ are 
uniquely fixed, while that for $\Psi^D_R$ is ambiguous. 
Once the KSVZ Yukawa interaction $S\overline{\Psi^D_L}\Psi_R^D$ is turned on, 
$\mathbf{B}$ and $\mathbf{L}$ for $\Psi^D_R$ are uniquely determined. \\

The $\mathbf{B+L}$ conserving operator $\mathcal{O}_7$ can be 
systematically constructed from the following 
$\mathbf{B}$ and $\mathbf{L}$ conserving operators, 
%
\begin{align}
{\mathcal O}\,'_6 
= \big\{ \Psi_R^Uu_R^{}d_R^{}e_R^{}, u_R^{}u_R^{}\Psi_R^De_R^{}, 
\Psi_R^Ud_R^{}QL, u_R^{}\Psi_R^DQL, \Psi_R^UQQe_R^{} \big\}, 
\end{align}
%
where one of a SM quark singlet $q_R^{}$ is substituted by 
$\Psi_R^q$ in $\widehat{\mathcal{O}}_6$. 
Similarly to the method developed with ${\mathcal O}\,'_\text{mix}$, 
${\mathcal O}\,'_6$ determines the baryon (lepton) number of 
KSVZ quarks as $\mathbf{B}(\Psi^q_R)=-2/3$ and $\mathbf{L}(\Psi^q_R)=-1$. 
If we take $\Psi_L^q$ from ${\mathcal O}\,'_\text{mix}$ and 
$\Psi_R^q$ from ${\mathcal O}\,'_6$, the KSVZ Yukawa interaction 
in Eq.~\eqref{Eq:KSVZ2} induces ${\mathcal O}_7$. 
The higher order operator ${\mathcal O}_8$ can also be derived 
by considering the operator ${\mathcal O}\,''_6=d_R^{}\Psi_R^{U}\Psi_R^{U}e_R^{}$, 
where $\mathbf{B}(\Psi^U_R)=-1/6$ and $\mathbf{L}(\Psi^U_R)=-1/2$ 
correspond to $\mathbf{B}(S)=\mathbf{L}(S)=+1/2$. 
\\

\begin{figure}[tb]
\centering 
\includegraphics[scale=1]{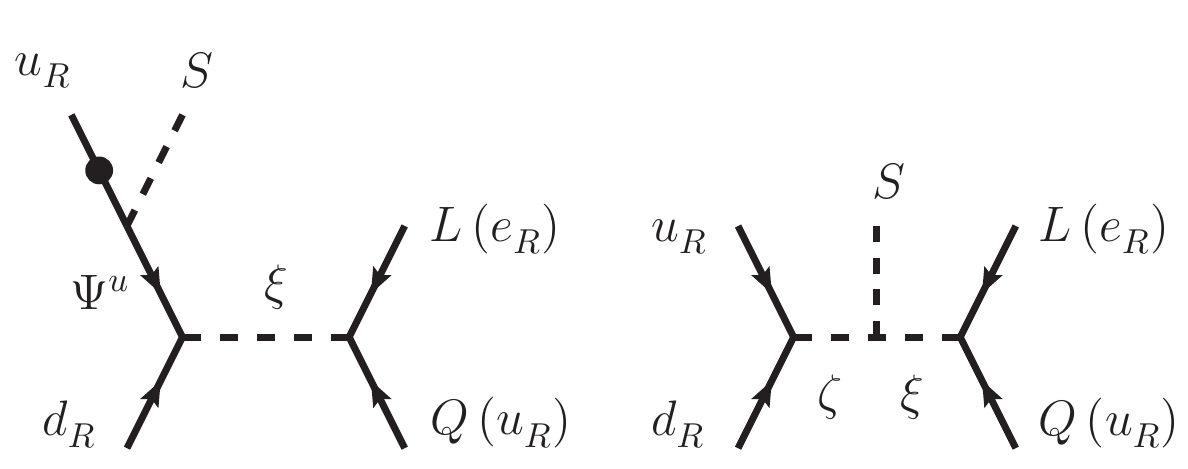}
\caption{Diagrams for ${\mathcal O}_7$ in a UV complete model.}
\label{Fig:B-L_axion}
\end{figure}

Let us show a UV complete example by introducing 
a singlet scalar $\xi$ with $Y=-1/3$. 
The leptoquark $\xi$ transforms as a fundamental representation of QCD, 
and the baryon and lepton numbers are assigned to $\mathbf{B}(\xi)=-1/3$ 
and $\mathbf{L}(\xi)=+1$. 
The baryon and lepton number conservation are assumed as PQ symmetry. 
Then, the dimension-six interaction of $\Psi_R^{U}d_R^{}QL$ is 
decomposed to the renormalizable interactions. 
The relevant Lagrangian is given by 
%
\begin{align}
{\mathcal L} 
=
&-y_\Psi^{}\, S\, \overline{\Psi^{Ua}_L} \Psi^{Ua}_R
-\mu_U^i\,\overline{\Psi^{Ua}_L}u_{iR}^{a}
-y_{\Psi D}^{i}\, \epsilon_{abc} \xi^a \overline{(\Psi_{R}^{Ub})^C} d_{iR}^{c}
\nonumber\\
& \qquad 
-\Big[ +y_{QL}^{ij} \overline{(Q_i^a)^C} (i\sigma_2) L_j
+y_{UE}^{ij}\,\overline{(u_{iR}^a)^C} e_{Rj}^{} \Big] (\xi^a)^* 
+ \text{H.c.}
\end{align}
%
We here show the color indices ($a, b, c=1, 2, 3$) and 
the flavor indices $(i,j=1,2,3)$ explicitly, 
and $\epsilon_{abc}$ is the Levi-Civita tensor. 
The Lagrangian is assumed to hold global $\mathbf{B}+\mathbf{L}$ symmetry 
as PQ symmetry.  
The operators of $QQ\,\xi$ and $u_R^{}d_R^{}\,\xi$, 
which lead dangerous dimension-six nucleon decays, 
are forbidden by $\mathbf{B}+\mathbf{L}$ symmetry unlike 
ordinary leptoquark models. 
Note that $\xi$ has definite $\mathbf{B}$ and $\mathbf{L}$ by construction, 
no diquark interaction is allowed. 
On the other hand, the model predicts nucleon decays via ${\mathcal O}_7$. 
The corresponding Feynman diagram is depicted in the left of 
Fig.~\ref{Fig:B-L_axion}. 
The low energy effective Lagrangian for the nucleon decay is derived as 
%
\begin{align}
{\mathcal L}_\text{eff}^{\Delta(\mathbf{ B}+\mathbf{L})=2}
= \epsilon_{abc}
\overline{(u_{iR}^a)^C} d_{jR}^{b}
\Big[
+C_{UDQL}^{ijkl} \big( \overline{(u_{kL}^c)^C} e_{lL}^{} 
-\overline{(d_{kL}^d)^C} \nu_{lL}^{} \big)
+C_{UDUE}^{ijkl}\, \overline{(u_{kR}^c)^C} e_{lR}^{}
\Big]
+\text{H.c.}
\end{align}
%
Here and hereafter, we omit the flavor indices. 
The Wilson coefficients are given by
%
\begin{align}
C_{UDQL}^{ijkl}
= -\frac{\mu_U^i y_{\Psi D}^{j} y_{QL}^{kl}}{M_\Psi M_\xi^2},\quad
C_{UDUE}^{ijkl}
= -\frac{\mu_U^i y_{\Psi D}^{j} y_{ue}^{kl}} {M_\Psi M_\xi^2}.
\end{align}
%
where mass of leptoquark is $M_\xi$, and that of KSVZ quark is 
$M_\Psi=y_\Psi^{}\langle S\rangle$.\footnote{
The typical mass scale of KSVZ quark can be as large as $\langle S \rangle$, 
while a TeV scale mass is possible by assuming the small Yukawa coupling constant 
$y_\Psi^{}\ll1$.} 
%
%
Using these effective interactions, the proton decay rates are 
%
\begin{align}
\Gamma_{p\to\pi^0e^+_i}
=&\frac{m_p}{32\pi} \bigg(1-\frac{m_\pi^2}{m_p^2}\bigg)^{\!\!2}\, 
\bigg( \Big| C_{UDQL}^{111i} \langle \pi^0|(ud)_R^{}u_L^{}|p\rangle_i \Big|^2
+\Big| C_{UDUE}^{111i} \langle \pi^0|(ud)_L^{}u_L^{}|p\rangle_i \Big|^2 \bigg),\\
\Gamma_{p\to\pi^+\overline{\nu}_i}
=&\frac{m_p}{32\pi} \bigg(1-\frac{m_\pi^2}{m_p^2}\bigg)^{\!\!2}\,
\Big| C_{UDQL}^{111i} \langle \pi^+|(du)_R^{}d_L^{}|p\rangle \Big|^2,\\
\Gamma_{p\to K^0e_i^+}
=&\frac{m_p}{32\pi} \bigg(1-\frac{m^2_K}{m^2_p}\bigg)^{\!\!2}\,
\bigg( \Big| C_{UDQL}^{121i} \langle K^0|(us)_R^{}u_L^{}|p\rangle_i \Big|^2
+\Big| C_{UDUE}^{121i} \langle K^0|(us)_L^{}u_L^{}|p\rangle_i \Big|^2 \bigg)\\
\Gamma_{p\to K^+\overline{\nu}_i}
=&\frac{m_p}{32\pi} \bigg(1-\frac{m_K^2}{m_p^2}\bigg)^{\!\!2}\, 
\Big| C_{UDQL}^{121i} \langle K^+|(us)_R^{}d_L^{}|p\rangle
+C_{UDQL}^{112i} \langle K^+|(ud)_R^{}s_L^{}|p\rangle\Big|^2,\\
\Gamma_{p\to\eta^0e^+_i}
=&\frac{m_p}{32\pi} \bigg(1-\frac{m_\eta^2}{m_p^2}\bigg)^{\!\!2}
\bigg( \Big| C_{UDQL}^{111i} \langle \eta^0|(ud)_R^{}u_L^{}|p\rangle_i \Big|^2
+\Big| C_{UDUE}^{111i} \langle \eta^0|(ud)_L^{}u_L^{}|p\rangle_i \Big|^2 \bigg).
\end{align}
%
where the nucleon matrix elements are taken from 
the lattice simulation result\cite{Ref:MEbyLAT}. 
%
%
For simplicity,  we assume $y_{QL}^{ij}=y_{ue}^{ij}=y_{q\ell}^{}$, 
$y_{\Psi D}^i=y_{\Psi D}^{}$ and $\mu_U^i=\mu_U^{}$ in numerical estimates. 
The proton (partial) lifetime is evaluated as functions of model parameters as
%
\begin{align}
\tau_{p\to\pi^0e^+}^{}
&\simeq (2.4\times10^{34}\,\text{yrs}) \times
\bigg( \frac{0.2}{\mu_U^{}/M_\Psi} \bigg)^{\!\!2} 
\bigg( \frac{1}{|y_{\Psi D}^{}|} \bigg)^{\!\!2} 
\bigg( \frac{1}{|y_{q\ell}^{}|} \bigg)^{\!\!2} 
\bigg( \frac{M_\xi}{2.0\times10^{15}\text{GeV}} \bigg)^{\!\!4},\\
\tau_{p\to\pi^0\mu^+}^{}
&\simeq (1.6\times10^{34}\,\text{yrs}) \times
\bigg( \frac{0.2}{\mu_U^{}/M_\Psi} \bigg)^{\!\!2} 
\bigg( \frac{1}{|y_{\Psi D}^{}|} \bigg)^{\!\!2} 
\bigg( \frac{1}{|y_{q\ell}^{}|} \bigg)^{\!\!2} 
\bigg( \frac{M_\xi}{1.7\times10^{15}\text{GeV}} \bigg)^{\!\!4},\\
\tau_{p\to\pi^+\overline{\nu}_i}^{}
& \simeq (3.9\times10^{32}\,\text{yrs}) \times
\bigg( \frac{0.2}{\mu_U^{}/M_\Psi} \bigg)^{\!\!2} 
\bigg( \frac{1}{|y_{\Psi D}^{}|} \bigg)^{\!\!2} 
\bigg( \frac{1}{|y_{q\ell}^{}|} \bigg)^{\!\!2} 
\bigg( \frac{M_\xi}{6.9\times10^{14}\text{GeV}} \bigg)^{\!\!4},\\
\tau_{p\to K^0e^+}^{}
&\simeq (1.0\times10^{33}\,\text{yrs}) \times
\bigg( \frac{0.2}{\mu_U^{}/M_\Psi} \bigg)^{\!\!2} 
\bigg( \frac{1}{|y_{\Psi D}^{}|} \bigg)^{\!\!2} 
\bigg( \frac{1}{|y_{q\ell}^{}|} \bigg)^{\!\!2} 
\bigg( \frac{M_\xi}{6.0\times10^{14}\text{GeV}} \bigg)^{\!\!4},\\
\tau_{p\to K^0\mu^+}^{}
&\simeq (1.6\times10^{33}\,\text{yrs}) \times
\bigg( \frac{0.2}{\mu_U^{}/M_\Psi} \bigg)^{\!\!2} 
\bigg( \frac{1}{|y_{\Psi D}^{}|} \bigg)^{\!\!2} 
\bigg( \frac{1}{|y_{q\ell}^{}|} \bigg)^{\!\!2} 
\bigg( \frac{M_\xi}{6.7\times10^{14}\text{GeV}} \bigg)^{\!\!4},\\
\tau_{p\to K^+\overline{\nu}_i}^{}
&\simeq (6.6\times10^{33}\,\text{yrs}) \times
\bigg( \frac{0.2}{\mu_U^{}/M_\Psi} \bigg)^{\!\!2} 
\bigg( \frac{1}{|y_{\Psi D}^{}|} \bigg)^{\!\!2} 
\bigg( \frac{1}{|y_{q\ell}^{}|} \bigg)^{\!\!2} 
\bigg( \frac{M_\xi}{1.2\times10^{15}\text{GeV}} \bigg)^{\!\!4},\\
\tau_{p\to\eta^0e^+}^{}
&\simeq (1.0\times10^{34}\,\text{yrs}) \times
\bigg( \frac{0.2}{\mu_U^{}/M_\Psi} \bigg)^{\!\!2} 
\bigg( \frac{1}{|y_{\Psi D}^{}|} \bigg)^{\!\!2} 
\bigg( \frac{1}{|y_{q\ell}^{}|} \bigg)^{\!\!2} 
\bigg( \frac{M_\xi}{1.0\times10^{15}\text{GeV}} \bigg)^{\!\!4},\\
\tau_{p\to\eta^0\mu^+}^{}
&\simeq (4.7\times10^{33}\,\text{yrs}) \times
\bigg( \frac{0.2}{\mu_U^{}/M_\Psi} \bigg)^{\!\!2} 
\bigg( \frac{1}{|y_{\Psi D}^{}|} \bigg)^{\!\!2} 
\bigg( \frac{1}{|y_{q\ell}^{}|} \bigg)^{\!\!2} 
\bigg( \frac{M_\xi}{8.1\times10^{14}\text{GeV}} \bigg)^{\!\!4}, 
\end{align}
%
where $\tau_{p\to M\ell_i}=\Gamma_{p\to M\ell_i}^{-1}$, 
$\mu_U^{}/M_\Psi$ characterizes the mixing between the SM and KSVZ quark. 
Among the current experimental bounds\cite{Ref:p->pie,Ref:p->piv,Ref:p->Ke,Ref:p->Kmu,Ref:p->Kv1,Ref:p->Kv2,Ref:p->etae} 
on the proton decay modes, we find that $p\to\pi^0e^+$ mode gives 
the strongest constraint on this model. 
%
%
In the above formula, we normalize the partial lifetime of the proton by their current lower bound, e.g. the lower bound for $p\to \pi^0e^+$ mode is 
$\tau_{p\to\pi^0e^+}^{} < 2.4\times10^{34}\,\text{yrs}$\cite{Ref:p->pie}. 
Among all the proton decay constraints, $p\to \pi^0e^+$ gives a strongest one. 
If we assume the order one Yukawa couplings 
$|y_{\Psi D}^{}| \sim |y_{q\ell}^{}| \sim 1$ and the order one mixing 
$\mu_U^{}/M_\Psi \sim 0.2$, the proton decay has 
already given a very stringent constraint on the new physics scale 
$M_\xi$ as $M_\xi \gtrsim 10^{15}\,$GeV, which is much larger than typical PQ scale 
of $10^{12-14}\,$GeV. 
We note that the scale of direct search of new colored particles 
at hadron colliders is only a few TeV, and those of indirect searches 
through the flavor changing observables are about $100$--$1000$ TeV 
by assuming the order one coupling constants. 
Therefore, the proton decay search is the most promising way 
to explore this model. 
Note that the other type of UV completion is also possible with 
more particles. 
For instance, if we introduce a QCD anti-fundamental scalar singlet diquark 
$\zeta$ with $Y=+1/3$ together with $\xi$, then the new Yukawa interaction 
$u_R^{}d_R^{}\zeta^*$ and the source of the mass mixing between 
the diquark and the leptoquark, $\zeta^*\xi^*S$, are allowed. 
The corresponding Feynman diagram for this UV completion is shown 
in the right of Fig.~\ref{Fig:B-L_axion} as an example. 

We may add right-handed neutrinos without imposing 
the lepton number symmetry in these setup. 
Since the Majorana mass term for right-handed neutrinos breaks 
the continuous lepton number symmetry explicitly, 
the baryon number symmetry is solely identified as PQ symmetry in this case. 
Thus, the `Sakharaxion' scenario is realized in a minimal way, 
where the Sakharon\cite{Ref:Sakharon}, 
pNGB from the spontaneous baryon number violation, is 
identified as an axion. 

Finally, we comment on the axion physics of the models. 
The model predictions for the axions are not changed from the minimal KSVZ model. 
This feature is common to the Majoraxion model.  
The constraints on the axion decay constant, e.g. from the duration time 
of neutrinos from the supernova SN 1987A\cite{Ref:Fa-SN1987A}, are applicable 
without modification. 
The possible axion dark matter scenario\cite{Ref:Axion-DM} can also be combined. 
Therefore, the model is examined only through the baryon number violation or 
the effect of new particles.

\subsection{Axion models based on $\Delta\mathbf{(B-L)}=2$ symmetry breaking}
The $d=7$ operators of $qqq\overline{\ell}\varphi$ violate 
the baryon and lepton number as $\Delta \mathbf{B}=-\Delta \mathbf{L}=1$. 
which are given by 
%
\begin{align}
\widehat{\mathcal O}_7 
= \big\{ u_R^{}d_R^{}d_R^{}\overline{L}H^*, d_R^{}d_R^{}d_R^{}\overline{L}H, 
d_R^{}d_R^{}Q\overline{e}_R^{}H^*, d_R^{}QQ\overline{L}H^* \big\}. 
\label{Eq:O7}
\end{align}
%
The global $\mathbf{B+L}$ symmetry is kept, 
while the $\mathbf{B-L}$ symmetry is not. 
The $\mathbf{B-L}$ symmetry may be restored with 
a complex scalar $S$ as 
%
\begin{align}
{\mathcal O}_8=S^*\,\widehat{\mathcal O}_7. 
\label{Eq:B-L}
\end{align}
%
The baryon and lepton numbers of $S$ are fixed to be 
$\mathbf{B}(S)=-\mathbf{L}(S)=+1$.  

As discussed in the previous subsection, 
the $\mathbf{B}$ and $\mathbf{L}$ charges for $\Psi_L\,(\Psi_R)$ 
can be fixed by ${\mathcal O}\,'_\text{mix}$. 
On the other hand, those for $\Psi_R\,(\Psi_L)$ is determined by 
the following $\mathbf{B}$ and $\mathbf{L}$ conserving operators, 
%
\begin{align}
{\mathcal O}\,'_7 
= \big\{ \Psi_R^Ud_R^{}d_R^{}\overline{L}H^*, 
u_R^{}\Psi_R^Dd_R^{}\overline{L}H^*, 
\Psi_R^Dd_R^{}d_R^{}\overline{L}H, 
\Psi_R^Dd_R^{}Q\overline{e}_R^{}H^*, 
\Psi_R^DQQ\overline{L}H^* \big\},  
\end{align}
%
where the SM singlet quarks are substituted by $\Psi_R$, and 
%
\begin{align}
{\mathcal O}\,'_6 
= \big\{ d_R^{}d_R^{}\Psi^D_L\overline{e}_R^{}, 
d_R^{}\Psi^D_LQ\overline{L} \big\}, 
\end{align}
%
where $QH^*$ is replaced by $\Psi^D_L$ in Eq.\eqref{Eq:O7}. 
Thus, we obtain $\mathbf{B}(\Psi^q_{L/R})=-2/3$ and 
$\mathbf{L}(\Psi^q_{L/R})=+1$ from these connections. 
The extensions to higher order operators are straightforward. 
\\

\begin{figure}[tb]
\centering 
\includegraphics[scale=1]{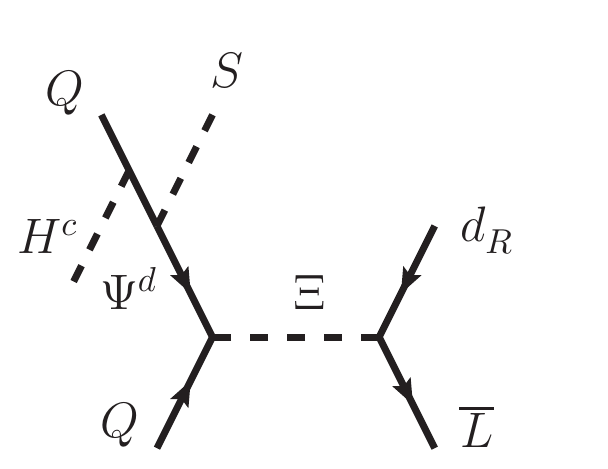}
\caption{Diagrams for ${\mathcal O}_8$ in a UV complete model.}
\label{Fig:B+L_axion}
\end{figure}

As an example, we give a UV completion of this type of models by introducing 
a doublet scalar $\Xi$ with $Y=-2/3$. 
The $\Xi$ transforms as a fundamental representation of QCD, 
and the baryon and lepton numbers are assigned to $\mathbf{B}(\Xi)=+2/3$ 
and $\mathbf{L}(\Xi)=-1$. 
The operator of $d_R^{}\Psi^D_LQ\overline{L}$ 
is decomposed by the following renormalizable interactions 
%
\begin{align}
{\mathcal L} 
=&-y_\Psi^{}\,S^*\overline{\Psi^{Da}_L}\Psi^{Da}_R 
-{{y'}_D^i}\overline{Q_i^a}H \Psi_R^{Da}
-y_{\Psi Q}^{i}\,\epsilon_{abc}\,\Xi^a\overline{(\Psi_{L}^{Db})^C} Q_{i}^{c}
-y_{\overline{L}D}^{ij} (\Xi^a)^* \overline{L_i}d_{jR}^a
+ \text{H.c.}
\end{align}
%
The global $\mathbf{B}-\mathbf{L}$ symmetry is imposed as PQ symmetry. 
This model predicts $\mathbf{B}+\mathbf{L}$ conserving nucleon decays 
via ${\mathcal O}_8$ as shown in Fig.~\ref{Fig:B+L_axion}. 
The low energy effective Lagrangian for the nucleon decay is derived as 
%
\begin{align}
{\mathcal L}_\text{eff}^{\Delta(\mathbf{ B}-\mathbf{L})=2}
= 
C_{DQ\overline{L}D}^{ijkl}
\epsilon_{abc} \overline{(d_{iL}^a)^C}
\big(u_{jL}^{b}\overline{\nu_{kL}}+d_{jL}^{b}\overline{e_{kL}}\big)d_{lR}^c
+\text{H.c.}
\end{align}
with 
\begin{align}
C_{DQ\overline{L}D}^{ijkl}
=-\frac{{{y_D^i}'}^* y_{\Psi Q}^{j} y_{\overline{L}D}^{kl}}{M_\Xi^2M_\Psi}
\frac{v_{\text{EW}}^{}}{\sqrt2}.
\end{align}
%
where $M_\Xi$ denotes the mass of $\Xi$, and 
$v_\text{EW}^{}$ is the Higgs VEV of electroweak symmetry breaking. 
%
%
In this model, no charged lepton mode of proton decay is induced. 
Only neutrino modes are available 
%
\begin{align}
\Gamma_{p\to\pi^+\nu_i}
=&\frac{m_p}{32\pi} \bigg(1-\frac{m_\pi^2}{m_p^2} \bigg)^{\!\!2}\, 
\Big| -C_{DQ\overline{L}D}^{11i1} \langle\pi^+|(ud)_R^{}d_L^{}|p\rangle \Big|^2, \\
\Gamma_{p\to K^+\nu_i}
=&\frac{m_p}{32\pi} \bigg(1-\frac{m_K^2}{m_p^2}\bigg)^{\!\!2}\,
\Big| -C_{DQ\overline{L}D}^{21i1} \langle K^+|(us)_R^{}d_L^{}|p\rangle
-C_{DQ\overline{L}D}^{11i2} \langle K^+|(ud)_R^{}s_L^{}|p\rangle \Big|^2, 
\end{align}
%
where the nucleon matrix elements are determined by the lattice calculation\cite{Ref:MEbyLAT}. 
%
%
We assume $y_{\overline{L}D}^{ij}=y_{\overline{L}D}^{}$, ${y_D^i}'=y_D'^{}$, 
$y_{\Psi Q}^i=y_{\Psi Q}^{}$ in the following numerical estimation for simplicity. 
The partial lifetimes calculated from inverse partial widths are   
%
\begin{align}
\tau_{p\to \pi^+\nu_i}
&\simeq (3.9\times10^{32}\,\text{yrs}) \times
\bigg( \frac{10^{-9}}{\tfrac{y_D'v_{\text{EW}}^{}}{\sqrt2}/M_\Psi} \bigg)^{\!\!2} 
\bigg( \frac{1}{|y_{\Psi Q}^{}|} \bigg)^{\!\!2} 
\bigg( \frac{1}{|y_{\overline{L}D}^{}|} \bigg)^{\!\!2} 
\bigg( \frac{M_\Xi}{4.9\times10^{10}\text{GeV}} \bigg)^{\!\!4},\\
\tau_{p\to K^+\nu_i}
&\simeq (6.6\times10^{33}\,\text{yrs}) \times
\bigg( \frac{10^{-9}}{\tfrac{y_D'v_{\text{EW}}^{}}{\sqrt2}/M_\Psi} \bigg)^{\!\!2} 
\bigg( \frac{1}{|y_{\Psi Q}^{}|} \bigg)^{\!\!2} 
\bigg( \frac{1}{|y_{\overline{L}D}^{}|} \bigg)^{\!\!2} 
\bigg( \frac{M_\Xi}{8.5\times10^{10}\text{GeV}} \bigg)^{\!\!4}, 
\end{align}
%
where $(\tfrac{y_D'v_{\text{EW}}^{}}{\sqrt2}/M_\Psi)$ characterizes 
the mixing between the SM and KSVZ quark.
Comparing the experimental bound of these two proton decay mode,  
$p\to K^+\bar{\nu}$ mode gives about an order of magnitude stronger bound 
than that on $p\to \pi^+\bar{\nu}$ mode. Thus, 
$p\to K^+\bar{\nu}$ mode is the most promising mode to explore this model. 

The impact of the Higgs field insertion of the operators appears 
in the mixing between the SM and KSVZ quark, which is strongly suppressed 
by $v_\text{EW}/M_\Psi$ 
Note that $v_\text{EW}^{}$ cannot be taken to be very large unlike the model 
discussed in the previous subsection. 
Even though the proton decay search has potential to probe high energy scale 
far beyond the collider reach 
if we assume the order one coupling constants. 

\subsection{More axion models}
At $d=7$, there is a dressed operator 
$LLHH|H|^2$ with $\Delta \mathbf{L}=2$. 
At $d=8$, we have $\Delta \mathbf{B}=\Delta \mathbf{L}=1$ 
operators\cite{Ref:BNV-LNV}
%
\begin{align}
\widehat{\mathcal O}_8 
= \big\{ u_R^{}u_R^{}QLH^cH^c, 
d_R^{}d_R^{}QLHH, 
d_R^{}QQe_R^{}HH, 
\widehat{\mathcal{O}}_6 |H|^2 \big\},  
\end{align}
%
where $\widehat{\mathcal{O}}_6$ expresses dimension-six $\mathbf{B+L}$ 
violating operators. 
These operators can also be used to construct an axion model. 
Since the model is controlled by the same symmetry, 
${\mathcal O}_7=S\widehat{\mathcal{O}}_6$ is allowed simultaneously  
with ${\mathcal O}_9=S\widehat{\mathcal{O}}_8$. 
The effects of $\mathcal{O}_7$ are dominated 
in the low energy phenomena such as nucleon decays. 

We can continue the same discussions with $d>8$ operators 
which contain lepton and baryon number violations. 
There are many variations of operators, which are characterized 
by $\Delta \mathbf{B}$ and $\Delta \mathbf{L}$ 
(see FIG.1 of \cite{Ref:BNV-LNV}). 
These operators can also be used to construct a model of axions. 
\\

%
\begin{figure}[tb]
\centering 
\includegraphics[scale=1]{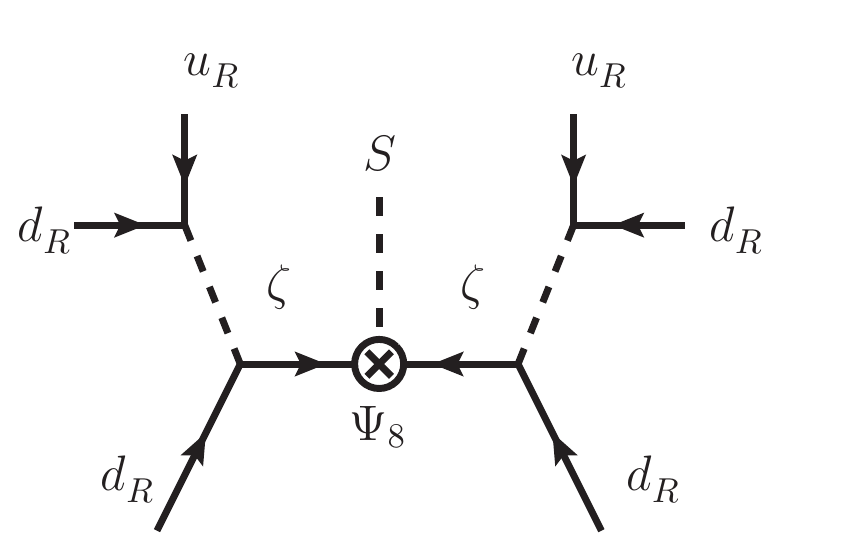}
\caption{Diagrams for $n$-$\bar{n}$ oscillation in a UV complete model.}
\label{Fig:nnbar-oscillation}
\end{figure}
%
\noindent\underline{A model for $n$-$\bar{n}$ oscillation}: 
At $d=9$, there are purely baryon number violating six-quark operators 
with $\Delta \mathbf{B}=2\,(\Delta \mathbf{L}=0)$. 
Therefore, the model predicts Sakharaxion without considering 
the lepton number violation. 
We here construct such a concrete model as an example. 
In order to build a model with minimal particle content, 
a color adjoint (Majorana) fermion $\Psi_8$ is introduced instead of 
the ordinary KSVZ quarks $\Psi$. 
Then, KSVZ Yukawa interaction is replaced by 
%
\begin{align}
{\mathcal L}_{\Psi_8} 
= -\frac12\,y_8^{}\,S^* \overline{(\Psi_{8L}^A)^C}\,\Psi_{8L}^A +\text{H.c.}
\end{align}
%
where $A=(1,2,\cdots,8)$.  
A model with this Yukawa interaction is known as 
the gluino-axion model\cite{Ref:GluinoAxion}, 
where the quantum charges of $\Psi_8$ are the same as {\it gluino} 
in the supersymmetric theories. 
In addition, a QCD color anti-fundamental weak singlet diquark $\zeta$ 
with $Y=+1/3$ is introduced. 
The baryon number symmetry is assumed as the PQ symmetry with 
$\mathbf{B}(\Psi_8)=+1$, $\mathbf{B}(\zeta)=+2/3$ and $\mathbf{B}(S)=+2$. 
The relevant Lagrangian is given by 
%
\begin{align}
{\mathcal L} 
=-y_\zeta^{ij} \epsilon_{abc} (\zeta_a)^* \overline{(u_{iR}^{b})^C} d_{jR}^c
-y_{8D}^i\, \zeta_a (T^A)_{~b}^{a}\,\overline{\Psi_{8L}^A} d_{iR}^b
+\text{H.c.}
\end{align}
%
where $(T^A)_{~b}^{a}$ is the generator of $SU(3)_C$. 
This model predicts $\Delta\mathbf{B}=2$ process such as 
$n$-$\bar{n}$ oscillation as shown in Fig.~\ref{Fig:nnbar-oscillation}. 
The low energy effective Lagrangian for the $n$-$\bar{n}$ oscillation is 
calculated as 
%
\begin{align}
{\mathcal L}_\text{eff}^{\Delta\mathbf{ B}=2}
=\, C_{UUDDDD}^{ijklmn}\, T^{AAS}_{[ab][cd]\{ef\}}
\overline{(u_{iR}^a)^C} d_{jR}^b
\overline{(u_{kR}^c)^C} d_{lR}^d
\overline{(d_{mR}^e)^C}d_{nR}^f +\text{H.c.}
\end{align}
with 
\begin{align}
C_{UUDDDD}^{ijklmn}
= \frac{y_\zeta^{ij} y_\zeta^{kl} y_{8D}^m y_{8D}^n}
{12M_\zeta^4 M_8}, \quad
T^{AAS}_{[ab][cd]\{ef\}}
=\epsilon_{abe} \epsilon_{cdf} +\epsilon_{abf}\epsilon_{cde},
\end{align}
%
where $M_\zeta$ and $M_8^{}$ denote the masses of $\zeta$ and $\Psi_8$. 
%
%
Using the result of 
the neutron-antineutiron matrix element\cite{Rinaldi:2018osy}, 
the $n$-$\bar{n}$ oscillation rate is estimated as 
%
\begin{align}
\Gamma_{n\overline{n}}
=&\frac{10^{-9}\, \text{s}^{-1}}{(700\,\text{TeV})^{-5}}
\Bigg|(-4.2)\times\frac14C_{UUDDDD}^{111111}\Bigg|. 
\end{align}
%
Replacing the Wilson coefficient by the model parameters, we found 
%
\begin{align}
\tau_{n\overline{n}} = {\Gamma_{n\overline{n}}}^{-1}
= (7\times10^8\, \text{s}) \times 
\bigg(\frac{M_\zeta}{400\,\text{TeV}}\bigg)^{\!\!4}
\bigg(\frac{M_8}{400\,\text{TeV}}\bigg)
\bigg(\frac1{|y_\zeta^{11}|} \bigg)^{\!\!2}
\bigg(\frac1{|y_{8d}^1|}\bigg)^{\!\!2}.
\end{align}
%
The current lower limit on neutron-antineutron oscillation is 
$\tau_{n\overline{n}} >4.7\times 10^8\,\text{s}$, which is 
given by Super-Kamiokande\cite{Abe:2020ywm}. 
Assuming the order one Yukawa coupling constants and 
the common new particle masses, the mass is constrained to be larger 
than $400\,$TeV. 
This process explores much higher new physics scale than that of LHC 
direct searches for new colored particles. 
Comparably strong bounds of $10^{2\text{--}4}$ TeV\cite{Ref:Isidori} 
may be obtained from neutral meson mixing, if the diquark $\zeta$ generates 
tree-level four fermion interactions with heavy quarks $s, c$ and $b$. 
If the diquark is assumed to interact only with the first generation quarks, 
we can avoid these flavor constraints since severe bounds come from $K$, $D$ 
and $B$ meson data.\footnote{
A comprehensive study for the diquark flavor structure is found, for example, 
in Ref.~\cite{Ref:diquark-flavor}.}
We also note that the model only predicts baryon number violations by two units, 
and thus no ordinary nucleon decay with $\Delta \mathbf{B}=1$ is induced. 
\\

%
\begin{figure}[tb]
\centering 
\includegraphics[scale=1]{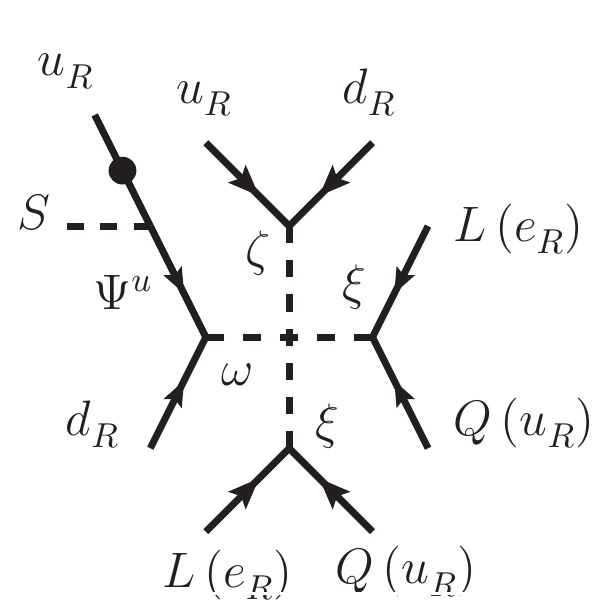}
\caption{Diagrams for di-nucleon decay in a UV complete model.}
\label{Fig:di-nucleon-decay}
\end{figure}
%
\noindent\underline{A model for di-nucleon decay}: 
Another interesting baryon number violating processes are di-nucleon decays, 
which are induced at $d=12$. 
Together with the QCD fundamental KSVZ quark $\Psi^U$ with $Y=+2/3$, 
one may introduce a leptoquark $\xi$, 
a diquark $\zeta$ 
and a tetraquark $\omega$ with $Y=-1/3$ and $\mathbf{L}(\omega)=+2$. 
The Lagrangian relevant to the di-nucleon decay is given by 
%
\begin{align}
{\mathcal L} 
=&-y_\Psi^{} S\,\overline{\Psi_L^{Ua}}\Psi_{R}^{Ua}
-\mu_U^i \overline{\Psi_L^{Ua}}u_{iR}^a
-\Big[ y_{QL}^{ij} \overline{(Q_{i}^a)^C} (i\sigma_2) L_j
+y_{UE}^{ij} \overline{(u_{iR}^a)^C} e_{jR}^{} \Big] (\xi^a)^* \nonumber \\
&-\epsilon_{abc} \Big[ y_{QQ}^{ij} \overline{(Q_{i}^b)^C} (i\sigma_2) Q_{j}^c
+y_{UD}^{ij} \overline{(u_{iR}^b)^C}d_{jR}^c \Big] (\zeta_a)^*
-y_{\Psi D}^i \epsilon_{abc} \overline{(\Psi_{R}^{Ua})^{C}}
d_{iR}^b\,\omega^c \nonumber \\
&-\lambda'\, \xi^a \xi^b \zeta_a (\omega^b)^* 
+\text{H.c.}
\end{align}
%
We note that $\mathbf{B}(\Psi^u_R)=-5/3$. 
The di-nucleon decay is generated by the diagram shown in Fig.~\ref{Fig:di-nucleon-decay}. 
The low energy effective Lagrangian for di-nucleon decays is written as 
%
\begin{align}
{\mathcal L}_\text{eff}^{\Delta(\mathbf{B}+\mathbf{L})=4}
= C_{UUUUDDEE}\, T_{[ab][cd]\{ef\}}^{AAS}
\overline{(u_{R}^a)^C}d_{R}^b\,
\overline{(u_{R}^c)^C}d_{R}^d\,
\overline{(u_{R}^e)^C}u_{R}^f\,
\overline{e_R^C}e_R^{}
+\text{H.c.}
\end{align}
with 
\begin{align}
C_{UUUUDDEE}
= -\frac{\lambda'\, \mu_U^{1}\, y_{\Psi D}^{1} y_{UD}^{11}(y_{UE}^{11})^2}
{4M_\omega^2 M_\zeta^2 M_\xi^4 M_\Psi},
\end{align}
%
where $M_\xi$, $M_\zeta$ and $M_\omega$ are masses of 
leptoquark, diquark and tetraquark, respectively. 
Let us evaluate the di-nuclaon decay rate with this effective interaction. 
Here, we set the parameters as $y_{QQ}=y_{QL}=0$ for simplicity. 
Following Ref.\cite{He:2021mrt}, the width of the di-nucleon decay is 
%
\begin{align}
\Gamma_{pp\to e^+e^+}\,
\simeq~&
\frac{(\text{TeV})^{16}}{2\times10^{26}\,\text{yrs}} 
\bigg( \frac{\rho_N^{}}{0.25\,\text{fm}^{-3}} \bigg)
\bigg( \frac{m_N^{}}{0.939\,\text{GeV}} \bigg)^{\!\!2}
\bigg( \frac{\Lambda_\text{QCD}^{}}{200\,\text{MeV}}\bigg)^{\!\!12}
\Big| C_{UUUUDDEE}\Big|^{2}. 
\end{align}
%
where $m_N=(m_p+m_n)/2$ is the mass of nucleon,
$\rho_N(\sim0.25\,\text{fm}^{-3})$ is the average nuclear matter density,
and $\Lambda_{\text{QCD}}$ is the QCD scale parameter.
In the present model, we found 
%
\begin{align}
\tau_{pp\to e^+e^+}
=\Gamma_{pp\to e^+e^+}^{-1}\,
\simeq~&
(5\times 10^{33}\,\text{yrs})
\times 
\bigg( \frac{M_\omega}{2\,\text{TeV}} \bigg)^{\!\!4}
\bigg( \frac{M_\zeta}{2\,\text{TeV}} \bigg)^{\!\!4}
\bigg( \frac{M_\xi}{2\,\text{TeV}} \bigg)^{\!\!8}
\nonumber \\
&\quad \times
\bigg( \frac{0.2}{\mu_U^1/M_\Psi} \bigg)^{\!\!2}
\bigg(\frac1{|\lambda'|}\bigg)^{\!\!2}
\bigg(\frac1{|y_{UD}^{11}|}\bigg)^{\!\!2}
\bigg(\frac1{|y_{UE}^{11}|}\bigg)^{\!\!4}
\bigg(\frac1{|y_{\Psi D}^{1}|}\bigg)^{\!\!2}. 
\end{align}
%
The lower limit of the lifetime of the $pp\to e^+e^+$ di-nucleon decay mode is 
$\tau_{pp\to e^+e^+}> 4.2\times10^{33}\,\text{yrs}$\cite{Sussman:2018ylo}. 
Even if we take order one new coupling constants, 
the common mass scale of new particles is about $2\,$TeV. 
This scale is almost the same as the current LHC bound on the colored new particles. 
Thus, the forthcoming high luminosity running of the LHC can help to test this model 
through the direct production of new colored particles. 
In order to avoid the constraint from the low energy flavor data, 
a specific flavor structure of the Yukawa coupling might be required. 
For example, if we assume that new colored particles solely couple 
the first generation fermions. 
In this case, effects on flavor changing decays of $\mu$, $\tau$ and mesons, 
and neutral meson mixing are forbidden at leading order. 
We also comment that no $\Delta\mathbf{B}=1$ proton decay as well as 
no $n$-$\bar{n}$ oscillation are predicted in this model. 
%

\section{Summary and discussions}
PQ symmetry is often introduced in order to solve strong CP problem
and is sometimes linked to other new physics scenarios such as 
lepton number violating neutrino masses. 
Possible connections between the PQ symmetry and the baryon number 
and lepton number symmetries have been studied in the extensions of KSVZ model. 
Since the KSVZ solves the strong CP problem in purely new physics sector, i.e., 
new KSVZ quark and a complex scalar, the baryon number and lepton number 
of new particles are undermined. 
In order to fix these quantum numbers, we have used  
the baryon and lepton number violating higher dimensional operators in the SM. 
Combining these operators with the scalar in the KSVZ model, 
we have developed the method to determine the baryon and 
lepton number of the new scalar. 
As a result, variants of the KSVZ axion model, which predict characteristic 
baryon number violations, are constructed. 
If we combine the scalar with $d=6\,(7)$ $\mathbf{B}+\mathbf{L}\,
(\mathbf{B}-\mathbf{L})$ violating operator, the axion model can also be 
explored through the nucleon decay experiment. 
With $d=9$ operator, the $n$-$\bar{n}$ oscillation 
is generated in the axion model. 
Since the VEV of the scalar violates the baryon number only by two units, 
$\Delta\mathbf{B}=1$ nucleon decay is forbidden in this model. 
Similarly with $d=12$ operator, the $\Delta(\mathbf{B}+\mathbf{L})=4$ 
di-nucleon decay is predicted while no other baryon number violation 
is generated from the lower dimension operators. 
The experimental search for the baryon number violation are expected to 
be upgrade in the near future\cite{Ref:JUNO,Ref:DUNE,Ref:HK}, 
it might be interesting to consider a diversity of baryon number violations 
other than the grand unified theories. 
%


\section*{Acknowledgments}
We would like to thank Ernest Ma for pointing out the key idea of this paper
and for careful reading of our manuscript. 
This work was supported by MEXT KAKENHI Grant Number JP18H05543. 


\end{document}